\documentclass[epj,referee]{svjour}
\usepackage{graphicx}
\usepackage{amssymb}
\begin{document}
\title{Non-conserved dynamics of steps on vicinal surfaces during electromigration-induced step bunching}
\author{M. Ivanov\inst{1} \and J. Krug\inst{1}
}                     
%
%
\institute{\inst{1} Institut f\"ur Theoretische Physik, Universit\"at zu K\"oln -  50937 K\"oln, Germany}
\date{Received: date / Revised version: date}
%
\abstract{We report new results on the non-conserved dynamics
of parallel steps on vicinal surfaces in the case of sublimation with 
electromigration and step-step interactions. The derived equations are valid in the quasistatic
approximation and in the limit $f^{-1}\gg l_D\gg l_{\pm} \gg l_i$, where $f$ is the inverse 
electromigration length, $l_D$ the diffusion length, $l_{\pm}$ the kinetic lengths and 
$l_i$ the terrace widths. The coupling between crystal sublimation and step-step interactions 
induces non-linear, non-conservative terms in the equations of motion. Depending on the initial conditions, 
this leads to interrupted coarsening, anticoarsening of step bunches or periodic switching between step trains
of different numbers of bunches.
} 
\titlerunning{Non-conserved dynamics during
  electromigation-induced step bunching}
\authorrunning{M. Ivanov and J. Krug}
\maketitle
\section{Introduction}
\label{intro}
For the theoretical study of homoepitaxial growth and sublimation of a crystal
in contact with the gas phase it is important to have a model, which includes the kinetic
processes and the different effects existing on the crystal surface. The classical model for 
the evolution of vicinal surfaces was introduced by Burton, Cabrera and Frank (BCF) \cite{Burton1951}.
It is based on the observation that the kink sites are those positions at the surface steps where 
the exchange between the adatom layer on the terraces and the solid phase takes place. On the mesoscopic scale
the change of the crystal volume is a result of the movement of the steps. On this scale we can reduce
a surface with straight steps to a one-dimensional step train. Such a surface may undergo \textit{step bunching}, an instability
where the steps move close to each other and form groups, called step bunches 
\cite{Jeong1999,Krug2005,PierreLouis2005,MisbahOlivierSaito}. 
   
The theoretical description of step bunching instabilities within the framework of the BCF-model and its extensions
has been the subject of much recent interest 
\cite{Sato1999,Sato2001,PierreLouis2003,KrugPimpinelli,popkovkrug,slava,Fok2007,Krug2010,IPK2010}. Here we focus
specifically on the effect of non-conservative processes on the non-linear evolution of a step train. As we reported
in \cite{IPK2010} for the problem of sublimation in the presence of Ehrlich-Schwoebel (ES) barriers \cite{Schwoebel1969,Ranguis}, 
non-conservative terms violating volume conservation in the co-moving frame arise generically from the 
interplay of sublimation and step-step interactions, and cause the interruption of the coarsening of the growing bunches or splitting
of a large bunch into several smaller bunches. In the present paper we expand this analysis to include the experimentally
relevant effect of surface electromigration \cite{Jeong1999,MisbahOlivierSaito,latyshev,Stoyanov1991,Stoyanov2011,Yang1996,Fujita1999,Yagi2001,Pelz,Ranguelov2008,Ranguelov2009,Usov2010,Usov2011}. In 1989, Latyshev
and collaborators discovered that by changing the direction of the direct heating current, a vicinal Si(111) surface switches between 
bunching and debunching \cite{latyshev}. Additionally, they observed
several distinct temperature regimes. In the so called regimes I and
III \cite{MisbahOlivierSaito} the bunching instability occurs only if the heating current is applied in the
down-step direction. On the other hand, for the same direction in
regime II debunching occurs, and bunching requires an up-step current. 
Here, we consider the first temperature regime, where the temperature is low enough in order to neglect step transparency (the motion of adatoms across steps) 
\cite{MisbahOlivierSaito,PierreLouis2003,Ranguelov2009}. 

Interrupted coarsening of electromigration-induced step
bunches in the presence of sublimation was previously observed
numerically by Sato and Uwaha \cite{Sato1999}, however a detailed analysis of the phenomenon was
not carried out due to the complexity of their model. Other
studies have approached the problem within the framework of weakly
nonlinear amplitude equations, which can be systematically derived
by an expansion around the instability threshold
\cite{MisbahOlivierSaito}. In this setting the non-conserved dynamics
is described on large scales by the Benney equation, which displays
either spatio-temporal chaos or an ordered array of bunches, but no
coarsening \cite{Sato1995,Misbah1996}. This macroscopic behavior is consistent
with the complex mesoscopic step dynamics revealed in the present
work. 

The paper is organized as follows. First, we sketch the derivation of the discrete step equations for the case of attachment-detachment limited kinetics and 
present the result of the linear stability analysis for very large wave lengths. 
Additionally, for comparison, we write down the corresponding equations for the case of growth. 
We then discuss the dicrete equations and their continuum limits for two special cases, where the kinetic asymmetry between ascending
and descending steps is caused solely by an ES-effect or by electromigration, respectively. Finally, we show the results of numerical simulations 
of the discrete step equations for the case with electromigration.

%
\begin{figure}
  \includegraphics{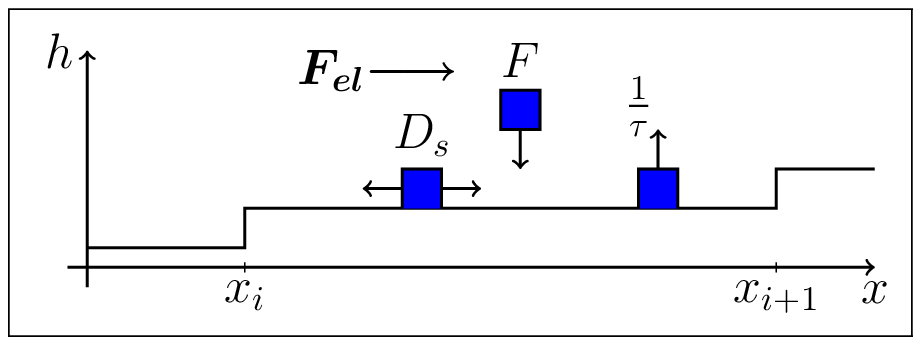}
\caption{(Color online) Sketch of the elementary processes in the
  Burton-Cabrera-Frank model.}
\label{fig.1}       
\end{figure}

\section{Model}

We consider an ascending one-dimensional step train with step edges located at positions $x_i$ (Fig.\ref{fig.1}). 
The starting point for the derivation of the equations of motion for the steps is 
the balance equation 
\begin{equation} \label{eq:BCFmodel}
\frac{\partial n_i}{\partial t}=D_s\left[\frac{\partial^2 n_i}{\partial x^2}-f\frac{\partial n_i}{\partial x}\right]-\frac{n_i}{\tau}+F\stackrel{!}{=}0
\end{equation}
for the concentration of adatoms $n_i(x,t)$ on the $i$-th terrace of width $l_{i}=x_{i+1}-x_{i}$. 
Here $D_s$ is the surface
diffusion constant, 
and $\tau$ is the average life time of an adatom before desorption. 
Together these two quantities define the diffusion length $l_D=\sqrt{D_s\tau}$, which sets
the scale of spatial variation for the adatom concentration.
The terrace is losing adatoms due to desorption at rate $1/\tau$ and gaining 
adatoms because of deposition with constant flux $F$. 
The applied direct heating force
$F_{el}$  causes a drift of adatoms for which we can use the Einstein relation and define a further length scale, 
the so called electromigration length $f^{-1}=k_BT/F_{el}$ \cite{Stoyanov1991,Sato1999}.
As is common in the field, we assume that the adatom concentration
adjusts instantaneously to the slowly moving steps, an
assumption that is know as the quasistatic 
approximation and amounts to setting $\partial_{t} n_i(x,t)=0$ in (\ref{eq:BCFmodel}). An
approach that goes beyond this approximation was recently presented by
Ranguelov and Stoyanov \cite{Ranguelov2008,Ranguelov2009}. 

 The general solution $n_i(x)$ of the ordinary differential equation (\ref{eq:BCFmodel}) can be specified using 
mass conservation at the steps as boundary conditions. A terrace of
width $l$ is bounded by two steps with positions $x=\pm l/2$, 
at which the flux continuity conditions 
\begin{eqnarray}\label{eq:bc}\nonumber
\frac{f_{-}}{\Omega}=D_s\left[\frac{\partial n}{\partial x}-fn\right]=+k_-[n-n_{eq}], \;    x=-\frac{l}{2},
\\
\frac{f_{+}}{\Omega}=D_s\left[\frac{\partial n}{\partial x}-fn\right]=-k_+[n-n_{eq}] \;    x=+\frac{l}{2},
\end{eqnarray}
must hold, where $\Omega$ is the cross section of an atomic site at
the step. 
The labels $+/-$ refer to quantities corresponding to the lower/upper terrace of a step.
The fluxes $f_{\pm}$ depend on both the difference of the adatom
concentration $n(x)$ compared to its equilibrium value  $n_{eq}$
and on the attachment/detachment to the steps with kinetic coefficients $k_{\pm}$. If the condition $k_{+}>k_{-}$
is fulfilled we speak about a standard ES effect
\cite{Schwoebel1969}. It induces an asymmetry in the concentration profiles $n_i(x)$ 
quantified by the asymmetry parameter 
\begin{equation}
\label{bES}
b_{ES} \equiv \frac{k_+-k_-}{k_++k_-} = \frac{l_--l_+}{l_-+l_+},
\end{equation}
where $l_{\pm}=D_s/k_{\pm}$ are called kinetic lengths. 

Apart from the attachment kinetics, a second effect incorporated into the boundary conditions (\ref{eq:bc}) is the step-step repulsion. The equilibrium
concentration $n_{eq}$ is determined by the chemical potential
$\Delta \mu_i$ at the $i$th step through the relation $n_{eq}\approx
n^0_{eq}(1+\triangle\mu_i/k_BT)$, and $\triangle\mu_i$ depends on the
widths of the two neighboring terraces $l_i$ and $l_{i-1}$ according to \cite{Jeong1999,grubermullins}
\begin{eqnarray}\label{eq:cp}
\frac{\triangle\mu_i}{k_BT}=-g\left(\frac{l^3}{l_{i}^3}-\frac{l^3}{l_{i-1}^3}\right)=\ :g\nu_i. 
\end{eqnarray}
where $l$ is the mean terrace spacing and $g$ is a
\textit{dimensionless} measure for the  strength of repulsion between
the steps \cite{Jeong1999,LiuWeeks}. 

\section{Step equations of motion}
Using Eqs.~(\ref{eq:BCFmodel},\ref{eq:bc},\ref{eq:cp}) we find the concentrations $n_i(x)$ 
for all terraces. The velocity of the $i-$th step is then given by the superposition of the fluxes coming from the
two neighboring terraces as $dx_i/dt=f_{-}+f_{+}$. 
Since the non-conservative terms of primary interest here arise from
sublimation, we discuss separately the limiting cases of pure
sublimation ($F=0, \; \frac{1}{\tau} > 0$) and pure growth ($F>0, \; \frac{1}{\tau} = 0$);
of course, in a typical experimental setup both processes may proceed simultaneously. 
For the case of pure sublimation and in the limit $f^{-1}\gg l_{D}$, we obtain the non-linear system
$$ R^{-1} \frac{dx_i}{dt}  =   
\frac{ \left[\left (\frac{l_+}{l_D^2}+\frac{f}{2}\right )s_i+\frac{1}{l_D}c_i\right] \gamma_i-\frac{1}{l_D}e^{-\frac{fl_{i}}{2}} \gamma_{i+1}}{\left[ \frac {f(l_--l_+)}{2}+1 \right]s_i +\frac {l_-+l_+} {l_D} c_i} $$
\begin{equation} \label{eq:fully}
+ \frac{ \left [(\frac{l_-}{l_D^2}-\frac{f}{2})s_{i-1}+\frac{1}{l_D}c_{i-1}\right ] \gamma_i -\frac{1}{l_D}e^{\frac{fl_{i-1}}{2}}\gamma_{i-1}}{\left [ \frac {f(l_--l_+)}{2}+1\right ]s_{i-1} + \frac {l_-+l_+} {l_D} c_{i-1}},
\end{equation}
where $R = n_{eq}^0 \Omega D_s$, $s_i = \sinh(l_i/l_D)$, $c_i = \cosh(l_i/l_D)$ and $\gamma_i = 1 + g \nu_i$.
The result (\ref{eq:fully}) contains all four length scales and
illustrates the complicated functional dependence for 
a simple one-dimensional step train.

To simplify these expressions we use the approximation of attachment-detachment limited
kinetics, $l_{D} \gg l_{\pm}\gg l $ \cite{LiuWeeks,Pimpinelli1994}. 
After some calculations along the lines of \cite{IPK2010} we arrive at
\begin{eqnarray}\label{eq:schwoebel}\nonumber
R_e^{-1} \frac{dx_i}{dt}&\approx&\gamma_i\left [  \frac{(1-b_{ES})}{2}l_{i}+\frac{(1+b_{ES})}{2}l_{i-1}\right ]\\\nonumber
&+&U\left (2\nu_{i}-\nu_{i+1}-\nu_{i-1} \right )\\
&-&\frac{b_{el}}{2} \left[\left(\gamma_{i}+\gamma_{i+1}\right)l_{i}-\left(\gamma_{i}+\gamma_{i-1}\right)l_{i-1} \right].
\end{eqnarray}
Here a second asymmetry parameter 
\begin{equation}
\label{bel}
b_{el} \equiv -\frac{fl_D^2}{l_-+l_+}
\end{equation}
incorporating the strength of electromigration
has been introduced, $R_e=(n^0_{eq}\Omega D_s)/l_D^2 = n_{eq}^0 \Omega/\tau$ is the constant rate with which the surface changes 
volume in a unit time (in the absence of non-linear, non-conservative terms, see below) and $U=(gl_D^2)/(l_-+l_+)$.

\section{Linear stability}
Equations similar to (\ref{eq:schwoebel}) can be derived when the surface is subject to a growth flux but
sublimation is absent ($F > 0, \; \frac{1}{\tau} = 0$). In that case the factor $\gamma_i$ in front of the 
square bracket on the right hand side of (\ref{eq:schwoebel}), which depends nonlinearly on the step
coordinates,  is replaced by the constant 
\begin{equation}
\label{eq:Gamma}
\gamma = -\frac{F\tau}{n^{0}_{eq}}.
\end{equation}
Analogous to the problem considered in \cite{IPK2010}, this implies qualitatively different instability conditions 
for growth and sublimation. Performing a standard linear stability analysis, in the limit
of large wavelength perturbations we find the instability conditions
\begin{eqnarray}\label{eq:cond}
b^{sub} &\equiv& 2b_{el}+b_{ES}>6g  \;\;\; \textrm{for} \; \textrm{sublimation}
\\ \label{eq:cond2}
b^{gr} &\equiv& 2b_{el}-\frac{F\tau}{n^{0}_{eq}}b_{ES}>0  \;\;\;
\textrm{for} \; \textrm{growth.}
\end{eqnarray}
In the case of growth step bunching merely requires the compound asymmetry parameter $b^{gr}$ to be positive, whereas for sublimation
the corresponding quantity $b^{sub}$ needs to exceed a positive threshold value $6g$.
This is an important consequence of the qualitatively different contributions to the balance eq.~(\ref{eq:BCFmodel}) that arise from desorption and deposition,
respectively. Note that in a general situation the instability  
conditions (\ref{eq:cond},\ref{eq:cond2}) can be combined into the form 
$b=(1-F\tau/n^{0}_{eq})b_{ES}+2b_{el}>6g$, which was already obtained in \cite{Fok2007}.

\section{Conservative and nonconservative dynamics}
Beyond the linear stability properties, a fundamental difference
between the scenarios of pure growth and sublimation is that
the surface dynamics is \textit{conservative} during growth but not during sublimation \cite{PierreLouis2003,IPK2010}. 
Here conservative dynamics implies that the rate of volume change of the crystal, obtained by summing the 
equations of motion over all steps $x_i$, is independent of the surface configuration \cite{Krug1997}. 
Indeed, replacing the $\gamma_i$ in front of the square brackets on the right hand side of (\ref{eq:schwoebel})
by the constant (\ref{eq:Gamma}) and summing over $i$, one readily obtains $\sum_i \dot x_i = - F \Omega L$,
where $L$ is the total length of the crystal. 

It is instructive to compare the structure of the non-conservative contributions induced 
during sublimation by the
configuration-dependent factors $\gamma_i$ in (\ref{eq:schwoebel}) for the two 
step bunching instabilities driven by electromigration and by an ES-effect, 
respectively. First we neglect the ES-effect, setting $b_{ES}=0$, which  
simplifies (\ref{eq:schwoebel}) into the form
\begin{eqnarray} \label{eq:equation}\nonumber
R_e^{-1}\frac{dx_i}{dt} &=& \frac{\gamma_i}{2}\left(l_{i}+l_{i-1}\right)+U\left (2\nu_{i}-\nu_{i+1}-\nu_{i-1} \right )\\
&-&b_{el}\left(l_{i}-l_{i-1}\right) \\ \nonumber 
&+&\frac{ g b_{el}}{2} \left(\nu_{i}l_{i-1}-\nu_{i}l_{i}  +\nu_{i-1}l_{i-1}-\nu_{i+1}l_{i} \right). 
\end{eqnarray}
The second group of terms on the RHS of eq.~(\ref{eq:equation}) with
prefactor $U$ arises from equilibrium step-step interactions and
stabilizes the regular step train, while the third group of terms describes
the effect of electromigration, which is  
stabilising or destabilising depending on the sign of $b_{el}$, i.e.,
the direction of $F_{el}$. The last group of terms arises from the
interplay of electromigration and step-step interactions. Since the
terms in this group cancel pairwise under summation over $i$, their
contribution is conservative and 
the only non-conservative contributions in eq.~(\ref{eq:equation})  
are the first terms multiplied by $\gamma_i$. 
 
For comparison, setting $b_{el} = 0$ eq.(\ref{eq:schwoebel}) reduces to the
equations derived in  \cite{IPK2010},
\begin{eqnarray} \label{eq:schwoebel2}\nonumber
& &R_e^{-1}\frac{dx_i}{dt}= \frac{\gamma_i}{2}\left(l_{i}+l_{i-1}\right)+U\left (2\nu_{i}-\nu_{i+1}-\nu_{i-1} \right )\\
&-&\frac{b_{ES}}{2}\left(l_{i}-l_{i-1}\right)+\frac{ g b_{ES}}{2} \left(\nu_{i}l_{i-1}-\nu_{i}l_{i}\right). 
\end{eqnarray} 
The difference between the two cases is that the terms proportional to
$gb_{el}$ on the RHS of eq.~(\ref{eq:schwoebel2}) do not cancel under summation with respect to
$i$, and thus are non-conservative. 
As will be shown in the next section, this gives rise to 
distinct contributions in the continuum limit. 

\section{Continuum equations} In previous work a systematic method
for deriving continuum equations of motion from the discrete step
dynamics was developed \cite{KrugPimpinelli,popkovkrug} which was applied to the model (\ref{eq:schwoebel2})
in \cite{IPK2010}. Briefly, the method can be seen as a kind of
Lagrange transformation \cite{JoachimKim} which replaces the
`Lagrangian' dynamics of particle-like steps by the 'Eulerian' evolution of  
the step density $m(x,t)$. The latter in turn defines a 
continuous height profile $h(x,t)$ through $m(x,t) = \frac{\partial h}{\partial x}$.

Here we wish to compare the two instability mechanisms described by
eqs.(\ref{eq:equation}) and (\ref{eq:schwoebel2}), respectively, on
the continuum level. Following the procedure outlined in
\cite{IPK2010} for both models, we find that the continuum evolution
equation takes the general form 
\begin{eqnarray} \label{eq:cont} \nonumber
& &\frac{\partial h}{\partial t}+\frac{\partial }{\partial
  x}\left[-\frac{3gm^{2}}{2}
  -\frac{m^{\prime}}{6m^{3}}+\frac{3U\left(m^2\right)^{\prime\prime}}{2m}
  - J_b\right] + 1= \\
&-&\frac{3g\left(m^{2}\right)^{\prime}}
{2}\left[ \frac{m^{\prime}
}{6m^{3}}\right]^{\prime} -\Phi_b,
\end{eqnarray}
where 
primes denote spatial derivatives. 
Here time $t$ is rescaled by $R_e$, length $x$ by the average step
distance $l$, and height $h$ is measured in units of the monoatomic
step height. The terms inside the square brackets on the LHS are
conservative, and the non-conservative contributions are collected on
the RHS of eq.(\ref{eq:cont}). The two models (\ref{eq:equation}) and
(\ref{eq:schwoebel2}) differ in the form of the contribution $J_b$ to
the conserved surface flux, and of the non-conservative term $\Phi_b$. 
Labeling the contributions due the ES-effect by \textit{ES} and those
due to electromigration by \textit{el}, respectively, the conservative
terms are given by 
\begin{equation}
J_b^{ES} = \frac{b_{ES}}{2m}, \;\;\;\;
J_b^{el} = \frac{b_{el}}{m}+3gb_{el}m^{\prime},
\end{equation}
and the non-conserved contributions are
\begin{equation} 
\Phi_b^{ES}=\frac{3gb_{ES}\left(m^{2}\right)^{\prime}}
{2}\left[\frac{1}{2m}\right]^{\prime}, \;\;\;\;
\Phi_b^{el} \equiv 0.
\end{equation}
As was discussed above, the terms in eq.(\ref{eq:equation})
proportional to 
$gb_{el}$ give rise to a conservative contribution, whereas the terms
in (\ref{eq:schwoebel2}) proportional to $g b_{ES}$ contribute to the non-conservative
part of the continuum equation.

In earlier work based on the continuum approach \cite{KrugPimpinelli,popkovkrug} the
non-conservative contributions were generally neglected because of the
smallness of $g$ \cite{IPK2010}, and it was therefore concluded that step bunching
phenomena induced by electromigration and by the ES-effect belong to
the same universality class \cite{KrugPimpinelli,Pimpinelli2002}.
However, it has subsequently become clear that small non-conservative terms may
qualitatively change the nonlinear dynamics of surface steps
\cite{IPK2010}, and the fact that these terms are of different form for the two
instability mechanisms implies that their equivalence needs to be
reexamined. In the following we therefore explore the nonlinear
behavior of the electromigration model (\ref{eq:equation}) using
numerical simulations.    

\begin{figure*}
\includegraphics[scale=0.7]{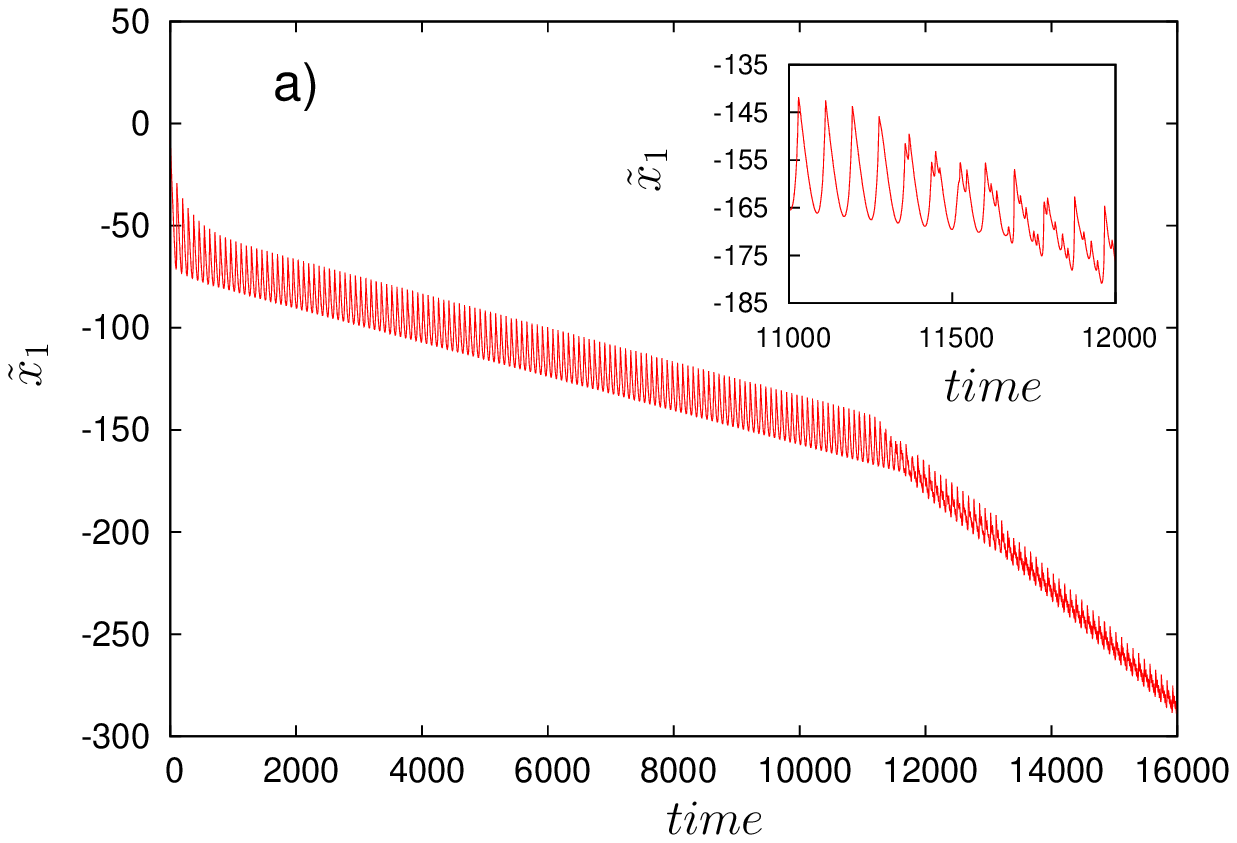}
\includegraphics[scale=0.7]{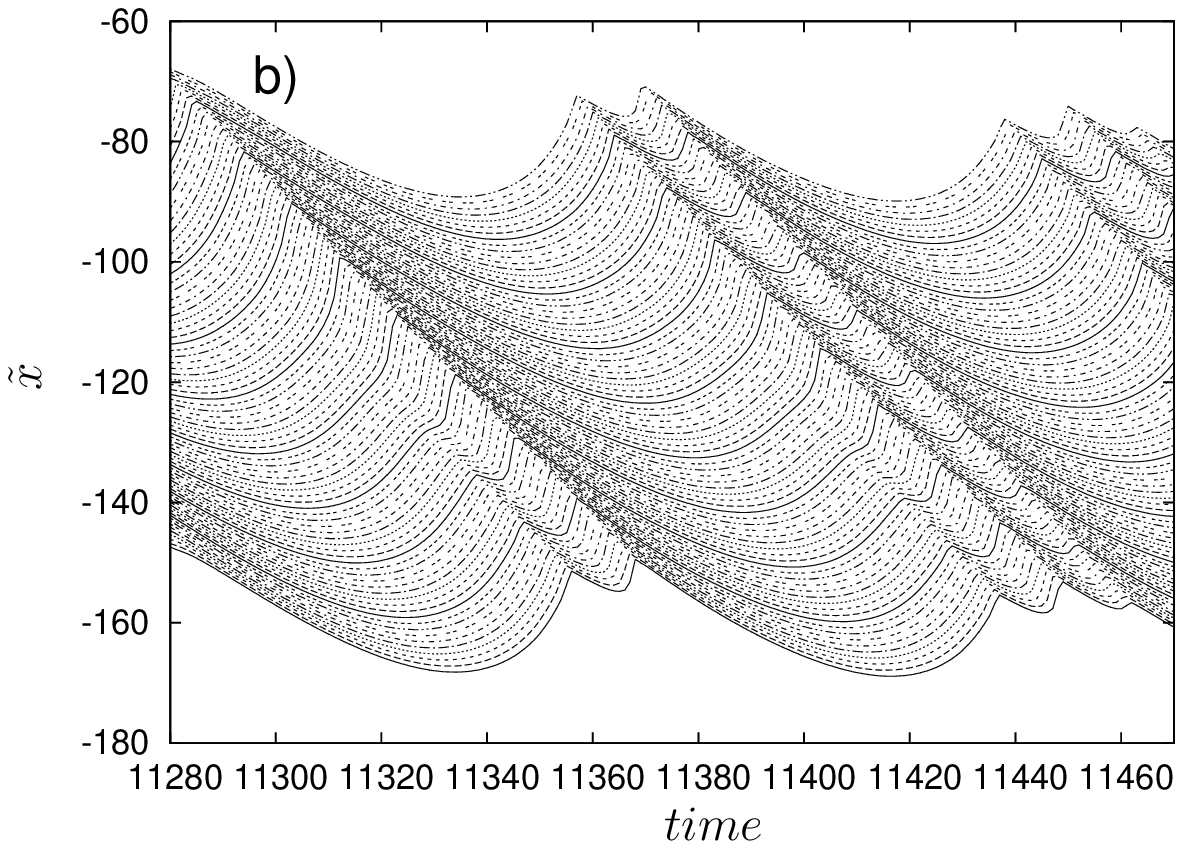} 
\includegraphics[scale=0.7]{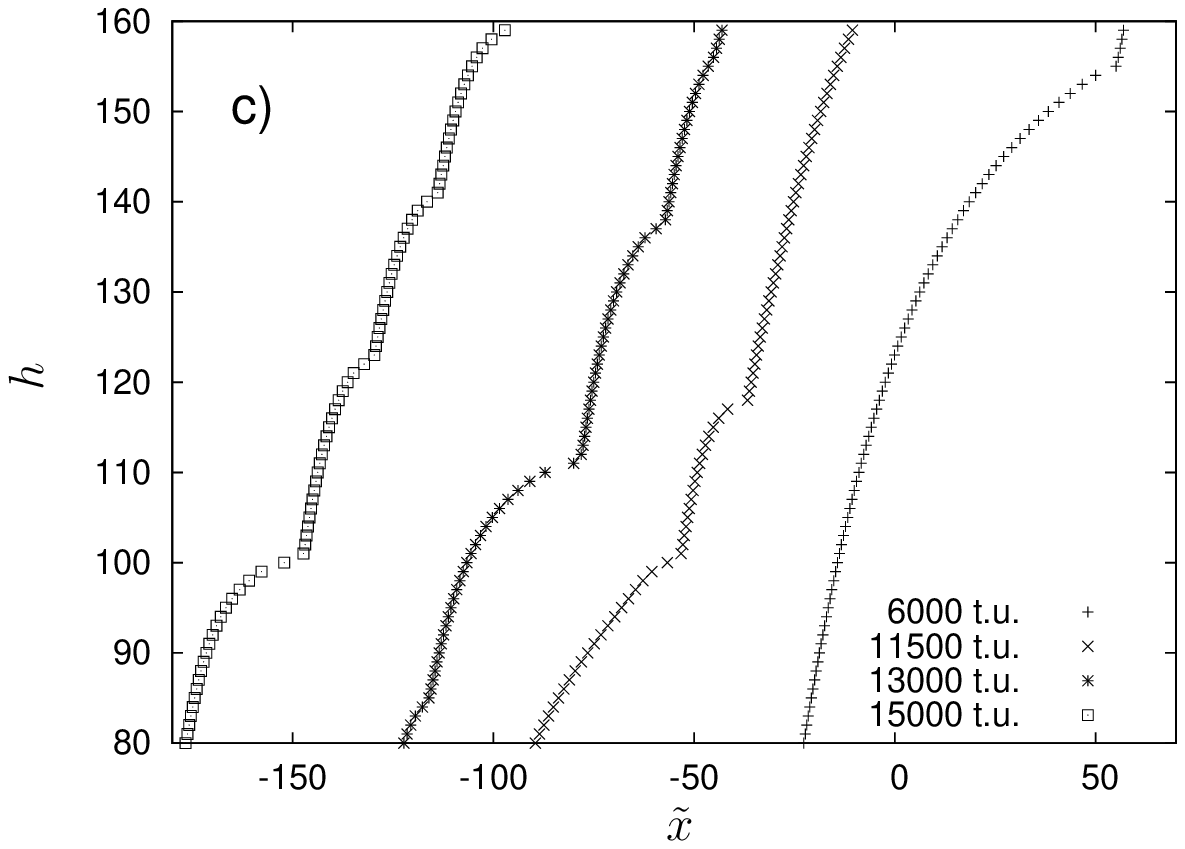}
\hspace*{.5cm} 
\includegraphics[scale=0.7]{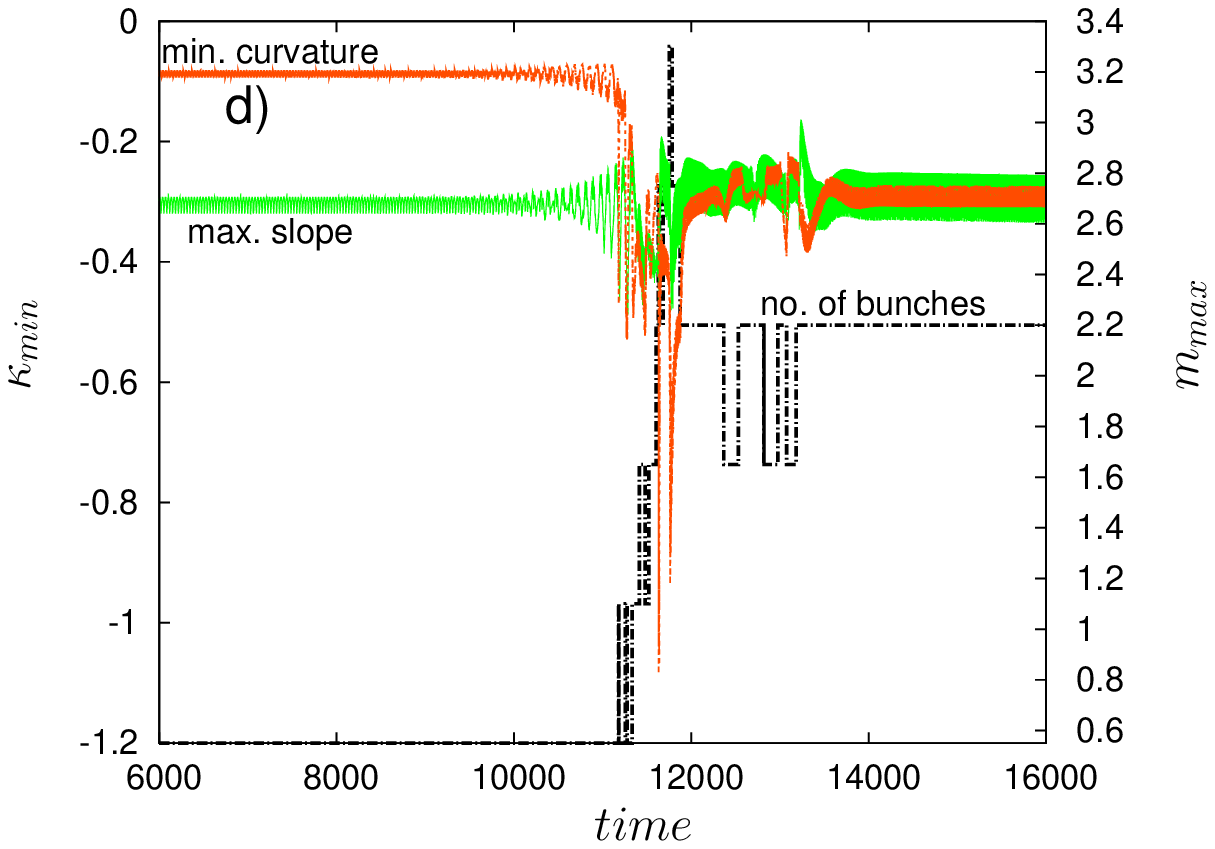} 
\caption{(Color online) An example for the splitting of a large bunch in a system of
  80 steps with parameters $b_{el}=0.4,\  g=0.05,\  U=0.05$. a) Time
  evolution of one of the steps. The inset shows a blowup around the
  onset of splitting. The period of oscillations prior to the
  breakup reflect the time required for the step to once traverse the (single)
  bunch. b) Plot of all step trajectories
  between $11280$ t.u. and $11470$ t.u. c) Comparison of the profiles
  after 6000 t.u., 11500 t.u., 13000 t.u. and 15000 t.u. d) Time
  evolution of the globally maximal slope, the globally minimal curvature and the number of bunches.}
\label{Fig_splitting}
\end{figure*}

\section{Nonlinear step dynamics}
Numerical simulations of eq.~(\ref{eq:equation}) were carried out  
using an odeint-type procedure \cite{Num} for systems of $M$ steps
with periodic boundary conditions. 
We consider the following ranges for the four independent parameters
of the model: $b_{el}\in[0,0.5]$, $U\in[0,0.5]$, $g\in[0,0.1]$
and $M<100$. Another degree of freedom is provided by the choice of
the initial condition. 
In general, we start the simulations with
two types of initial step train configurations: either a randomly
disturbed equidistant step train, or an
initial shock of closely spaced steps and a single large terrace. 
Step trajectores are shown in the co-moving
coordinate system $\tilde{x}_i(t)=x_i(t)-lt$, and we
normalize both the height of the (monoatomic) steps and the average terrace width $l$ to unity. The
time $t$ is rescaled by $R_e$ and we measure the integration time in time units (t.u.).
 For the description of the bunch geometry we use two measures: the maximal slope $m_{max} \equiv \max_{i} \{m_i \}$
and the minimal curvature $\kappa_{min} \equiv \min_{i} \{\kappa_i \} $, where $m_i=1/l_i$ and 
$\kappa_i=-8(l_{i+1}-l_i)/(l_{i+1}+l_i)^3$ respectively. A step is
defined to belong to a bunch, if its distance to the next closest 
step of the bunch is smaller than $l=1$.

\begin{figure*}
\includegraphics[scale=0.7]{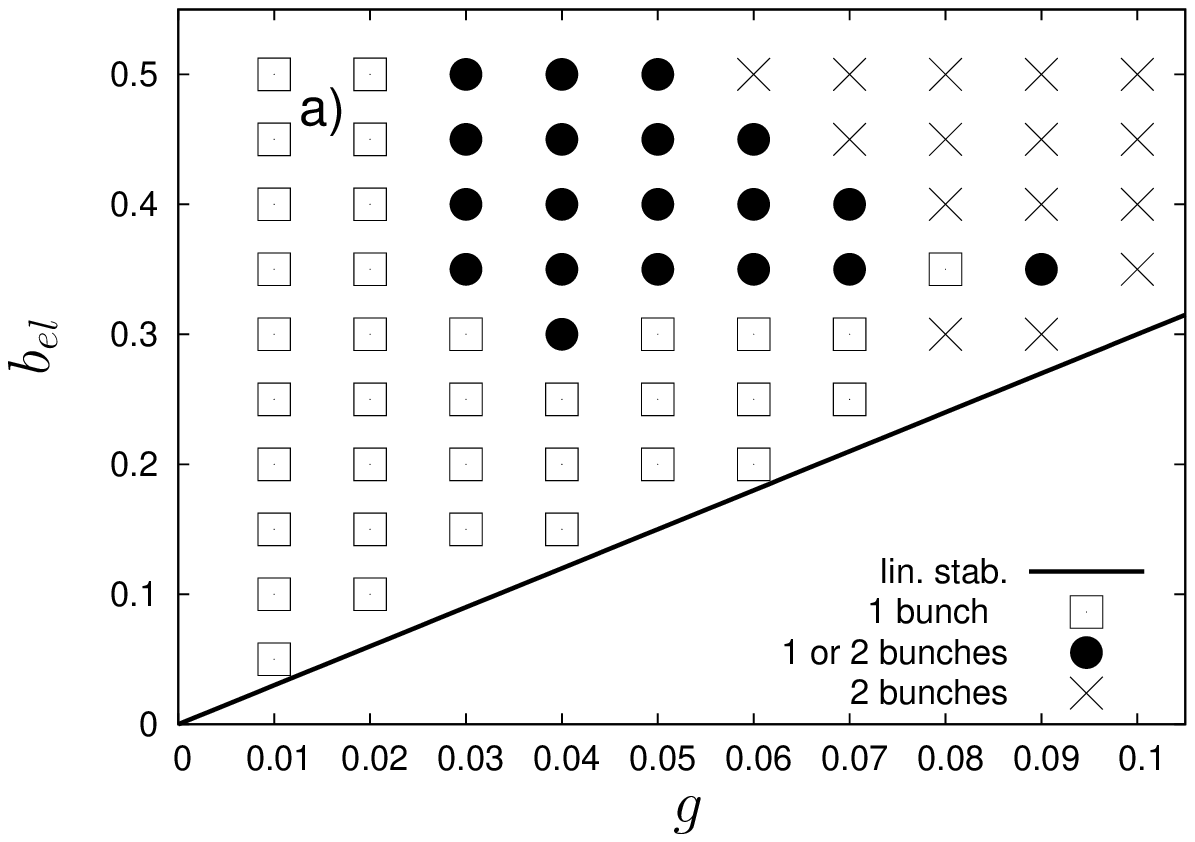}
\includegraphics[scale=0.7]{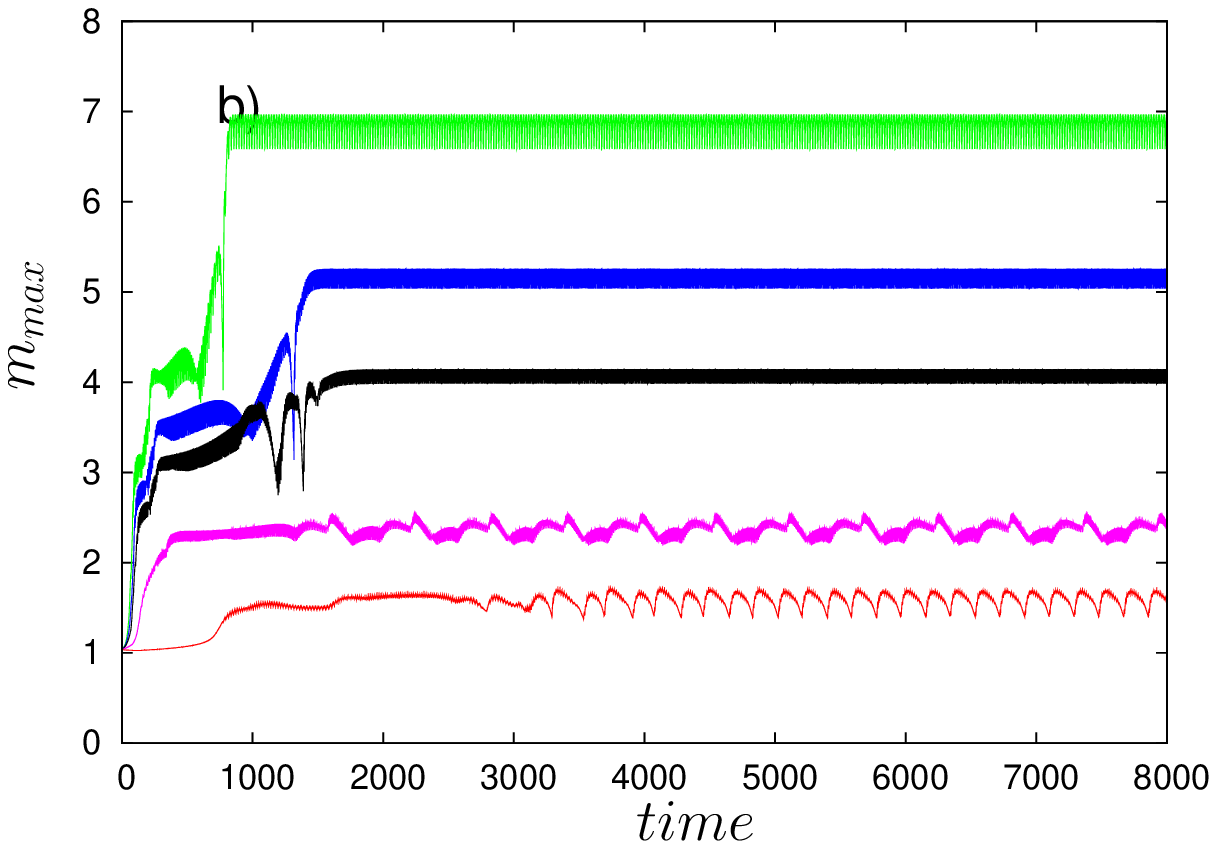}
\caption{(Color online) System with $U$=0.2, $M$=40 and fluctuating
  initial condition. a) Stability/instability diagram, showing the
  number of bunches in the final state, for different combinations of
  $b_{el}$ and $g$ - 
$\Box$: 1 bunch, $\bullet$: 1 or 2 bunches, $\times$: 2 bunches, below the line $b_{el}=3g$: stability. b) Time evolution of the maximal slope $m_{max}$ with $b_{el}$=0.35 and (from top to bottom) $g$=0.00, 0.01, 0.02, 0.05, and 0.09.} %
\label{Fig_splitting-fluct}%
\end{figure*}

An important consequence of the non-conservative character of the
dynamics is the phenomenon of \textit{anti-coarsening}, where an
initial large step bunch splits into smaller bunches \cite{IPK2010}. 
In fig.~\ref{Fig_splitting} we show an example of this behavior for
parameter values $b_{el}=0.4,\  g=0.05,\  U=0.05$, and $M=80$.
Figure~\ref{Fig_splitting}a) displays the movement of 
the first step in the train. After 11180 t.u. the large bunch splits for the first time into a large and a very small bunch, which is reflected  
in a clear change in the velocity of the step; smaller bunches move faster. Focusing on the splitting region one finds 
that the initial small bunch disappears again after 4 t.u., an event which is repeated at 11258 t.u. with a small bunch life time of 19 t.u.
In fig.~\ref{Fig_splitting}b) we show the trajectories of all steps in the train in the time window $11280-11470$ t.u. At 11333 t.u. the small bunch 
appears for the third time and at 11415 t.u. a third bunch arises for the first time. The switching between different numbers of bunches 
continues until the step train relaxes into four bunches, as seen in the height profiles in fig.~\ref{Fig_splitting}c).	
Figure~\ref{Fig_splitting}d) shows the corresponding time evolution of the minimal curvature $\kappa_{min}$ and the maximal 
slope $m_{max}$, along with the number of bunches. Both $\kappa_{min}$ and $m_{max}$  show clear changes in the
region of the splitting; however, whereas the maximal slope remains essentially the same after the splitting event, 
there is a significant decrease in the minimal curvature.

In fig.~\ref{Fig_splitting-fluct} we summarize results obtained from simulations starting from a randomly disturbed equidistant step array of 40 steps.
Figure~\ref{Fig_splitting-fluct}a) shows the phase diagram of the system in the $g$-$b_{el}$ plane at fixed $U=0.2$. 
Below the line $b_{el}=3g$ the system is linearly stable and
$m_{max}=m(x) \equiv 1$. Above this line we 
see three qualitatively different types of long-time behavior: steady solutions with one single bunch, with two bunches, and time-dependent solutions 
that switch periodically between one and two bunches.
In fig.~\ref{Fig_splitting-fluct}b) we plot the time evolution of the maximal slope $m_{max}$ at five points along the line $b_{el}$=0.35.  
For $g$=0 we see the usual coarsening behavior, in which the number of
bunches decreases in a step-wise fashion until a single bunch configuration is reached
and the system relaxes to a stationary
periodic state with a clearly bounded maximal slope. The remaining
temporal periodicity of $m_{max}$ is due to the permanent step exchange between the front and back end of the 
bunch (see also inset in fig.\ref{Fig_splitting}a)). Increasing $g$ the maximal slope decreases (while still maintaining the single bunch configuration), until at $g=0.05$ the regime of periodic switching
is reached, leading to a complex periodic pattern in $m_{max}$. 

\begin{figure}
\includegraphics[scale=0.7]{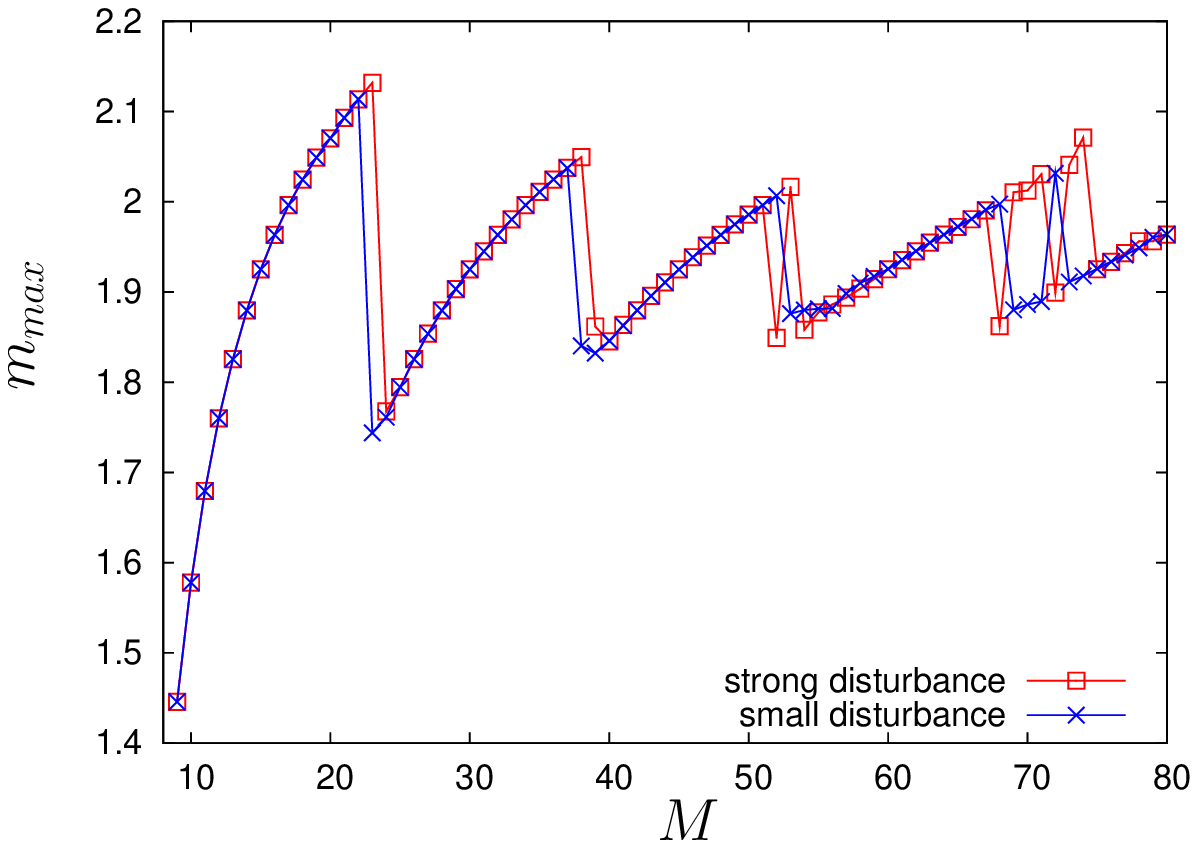}
\caption{(Color online) Dependence of the globally maximal slope
  $m_{max}$ on the number of steps $M$ for $g=U=0.04$, $b_{el}=0.2$
  and for large and small noise amplitude in the random initial condition.}%
\label{Fig_M_change}%
\end{figure}

Finally, in fig.~\ref{Fig_M_change} we plot the behavior of the maximal slope as a function of the number of steps for two different amplitudes of the initial disturbance, $g$=$U$=0.04, and 
$b_{el}$=0.2. Here $m_{max}$ is the global maximal slope, measured as its largest value for the last 500 t.u. of the simulation. We see that $m_{max}$ generally increases with $M$, 
but this behavior is interrupted by downward jumps every time the number of bunches that can fit into the system increases by one. This shows that the existence of stationary solutions
with multiple bunches can be seen as a consequence of the fact that, in the presence of non-conservative processes, the maximal slope is bounded from above \cite{IPK2010}. 
Near the transition between different numbers of bunches the system behavior depends very sensitively on the amplitude of fluctuations in the initial configuration, an 
effect that is particularly pronounced around $M=70$. 

\section{Conclusion}
In this work we have extended the non-conservative step bunching model presented in \cite{IPK2010} to include the effect of electromigration. The model applies to the first of the experimentally
observed temperature regimes on the Si(111) surface, where step transparency can be neglected. The general step equations of motion incorporating sublimation, the Ehrlich-Schwoebel effect, electromigration and 
step-step interactions were derived from the classical BCF model in the quasistatic approximation. For the case of attachment-detachment limited kinetics we compared the equations for growth and sublimation. In previous publications \cite{KrugPimpinelli,popkovkrug,slava,LiuWeeks} non-conservative contributions were neglected, because of the experimentally small prefactor $g$ \cite{Jeong1999,IPK2010}. 
Those terms were now taken into account and some important consequences were identified. First, on the level of linear stability analysis, they shift the instability condition on the dimensionless asymmetry parameter
$b$ by  $6g$, as was first pointed out in \cite{Fok2007}. This shift is present in the case of sublimation, but not in the case of growth \cite{IPK2010}. 
Moreover, in the case of sublimation the structure of the non-conservative terms differs depending on the underlying mechanism inducing the asymmetry between
ascending and descending steps. This leads to different continuum equations for step bunching caused by an ES-effect or by electromigration, respectively. 

Nevertheless, 
the numerical integration of the discrete step equations for the case with sublimation and electromigration reproduces qualitatively the results of \cite{IPK2010}. The non-linear, 
non-conservative terms supply a richness of dynamical behaviors in
this simple one-dimensional step model. There are steady solutions which contain more than one bunch, periodic switching 
between step trains of different numbers of bunches, and a sensitive dependence on the initial condition. This shows that the notion of universality between different types
of step bunching mechanisms, which was originally formulated on the basis of conservative continuum equations \cite{KrugPimpinelli,Pimpinelli2002}, 
can be applied also in the presence of non-conservative dynamics. 

In previous work on the conservative version of (\ref{eq:schwoebel}) a dynamical phase transition was identified which separates two qualitatively different regimes of 
step bunching distinguished by the presence or absence of crossing steps between bunches \cite{slava}. In our units this transition occurs at $b_{el}=1/2$, and experimental evidence
for its existence in the Si(111) system has recently been reported \cite{Usov2010}.  
In order to clearly bring out the effects due to the non-conservative nature of the dynamics, in the present study we have restricted ourselves to the 
parameter range $b_{el}\in[0,0.5]$, but the influence of non-conservative terms on the phase transition reported in \cite{slava} is clearly an interesting
topic for future work.

We thank V. Popkov for useful discussions.

\end{document}